\documentstyle[epsfig]{elsart}
\begin{document}

\newcommand{\re}{\mathop{\mathrm{Re}}}
\newcommand{\im}{\mathop{\mathrm{Im}}}
\newcommand{\D}{\mathop{\mathrm{d}}}
\newcommand{\I}{\mathop{\mathrm{i}}}
\newcommand{\E}{\mathop{\mathrm{e}}}

\begin{frontmatter}

\journal{Optics Communications}
%\date{}

\title{
Scheme for Attophysics Experiments at a  X-ray SASE FEL}

\author[DESY]{E.L.~Saldin}, 
\author[DESY]{E.A.~Schneidmiller},
and 
\author[Dubna]{M.V.~Yurkov}

\address[DESY]{Deutsches Elektronen-Synchrotron (DESY), 
Hamburg, Germany} 
\address[Dubna]{Joint Institute for Nuclear Research, Dubna, 
141980 Moscow Region, Russia}

\begin{abstract}

We propose a concept for production of high power coherent attosecond 
pulses in X-ray range. An approach is based on generation of 8th 
harmonic of radiation in a multistage HGHG FEL (high gain high harmonic 
free electron laser) configuration starting from shot noise.  
Single-spike phenomena occurs when electron bunch is passed through the 
sequence of four relatively short undulators. The first stage is a 
conventional "long" wavelength (0.8 nm)  SASE FEL which operates in the 
high-gain linear regime.  The 0.1 nm wavelength range is reached by 
successive multiplication (0.8 nm $\to$ 0.4 nm $\to$ 0.2 nm $\to$ 0.1 
nm) in a stage sequence.  Our study shows that the statistical 
properties of the high-harmonic radiation from the SASE FEL, operating 
in linear regime, can be used for selection of radiation pulses with a 
single spike in time domain.  The duration of the spikes is in 
attosecond range.  Selection of single-spike high-harmonic pulses is 
achieved by using a special trigger in data acquisition system.  The 
potential of X-ray SASE FEL at TESLA at DESY for generating attosecond 
pulses is demonstrated.  Since the design of XFEL laboratory at TESLA 
is based on the use of long SASE undulators with tunable gap, no 
special place nor additional FEL undulators are required for 
attophysics experiments.  The use of a 10 GW-level attosecond X-ray 
pulses at X-ray SASE FEL facility will enable us to track processes 
inside atoms.
  
\end{abstract}

\end{frontmatter}

\clearpage
\setcounter{page}{1}

\section{Introduction}

A general objective in the development of synchrotron radiation sources 
is to produce radiation that is brighter than that from existing 
sources, or to produce radiation that comes in shorter pulses.  
Significant progress in both of these directions has been reported 
recently by the TESLA collaboration. The results have been obtained at 
the TESLA Test Facility (TTF) at DESY \cite{ay}, using radiation pulses 
of 100 nm wavelength with sub-100 femtosecond pulse duration and peak 
power of approximately one GW. Comparing to present day synchrotron 
radiation sources its spectral brightness is more than a 100 million 
times higher, the radiation has full transverse coherence and pulse 
duration is reduced from the 100 picoseconds down to 100 femtoseconds 
in time domain. These demonstrations have been made possible by the 
technique of Self Amplified Spontaneous Emission Free Electron Laser 
(SASE FEL).  The generation of radiation in a linac driven SASE FEL has 
much a common with the generation of radiation in synchrotron source, 
the main difference in being the power dependence on number of 
electrons $N$.  The radiation generated in synchrotron sources is based 
on the spontaneous radiation of many electrons uncorrelated in space 
and time.  As a consequence, the radiation power scales linearly with 
the number of electrons in the bunch.  In the SASE FEL technique in 
order to increase the power and coherence of radiation, one has to 
force the electrons to emit coherently by compressing them into volume 
small compared to the wavelength of radiation. With complete 
micro-bunching, all electrons radiate almost in phase. This leads to a 
radiation power proportional to $N^2$ and thus amplification of many 
orders of magnitude with respect to the spontaneous emission.  TTF FEL 
shows that a laser for X-rays can be built on the basis of the SASE FEL 
principle. Technical design studies were presented for XFEL laboratory 
\cite{tdr} which should provide X-rays at wavelengths down to 0.1 nm in 
pulses of 100 fs duration. Peak spectral brightness would exceed those 
of synchrotron sources by over ten orders of magnitude.  Furthermore 
soft X-ray SASE FEL project with wavelengths down to 6 nm was started 
at DESY.  Commissioning of this facility will start in the year 2003 
\cite{ko}. 

The discussion in the scientific community over the past decade has 
produced many ideas for novel applications of the X-ray laser.  
Brilliance, coherence, and timing down to the femtosecond regime are 
the three properties which have the highest potential for new science 
to be explored with an XFEL. It is obvious that studies of time 
dependent phenomena can be tackled for the first time which relate the 
structural aspects  with the transition states of those electrons which 
are responsible for the formation process of intra-molecular bonds, 
clusters, nanoparticles, liquids, solids and hot dense plasmas.  
Femtosecond-resolution experiments with X-rays can possibly show us 
directly how matter is formed out of atoms. In fact, X-ray pulse 
duration even shorter than femtosecond may be useful for many 
scientific applications. A first question that arises is: why do 
researchers want laser-like X-ray radiation sources with shorter pulse 
durations? The reason is that phenomena inside atoms occur on 
sub-femtosecond timescale. Generating single attosecond X-ray pulses is 
one of the biggest challenges in physics.  The use of such a tool will 
enable us to trace processes inside atoms.  If there 
is any place where we have a chance to test the main principles of 
quantum mechanics in the pure way, this is it. 
 
\begin{figure}[tb]
\begin{center}
\epsfig{file=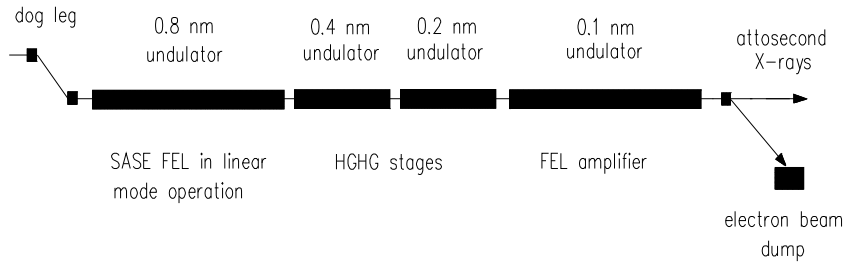,width=0.8\textwidth}
\end{center}
\caption{
Concept of the attosecond X-ray facility. XFEL
produces ultrafast X-ray pulses during a single pass of electron beam 
through a sequence of undulators which are resonant at different 
wavelengths.  The amplification process develops from shot noise. The 
single-spike pulse selection can be achieved by using special trigger 
in data acquisition system} 
\label{fig:pp31} 
\end{figure}

Our studies have shown that the X-ray SASE FEL holds a great promise as 
a source of radiation for generating high power, single attosecond 
pulses.  Two major developments have made this possible. It was shown
recently that the statistical properties of the high-harmonic 
radiation from the SASE FEL, operating in linear regime, can be used 
for selection of radiation pulses with a single spike in time domain 
\cite{tr}. In the case of X-ray FEL the duration of the spikes is in 
attosecond range. Selection of single-spike high-harmonic pulses can 
be achieved by using a special trigger in data acquisition system. 

\begin{figure}[tb]
\begin{center}
\epsfig{file=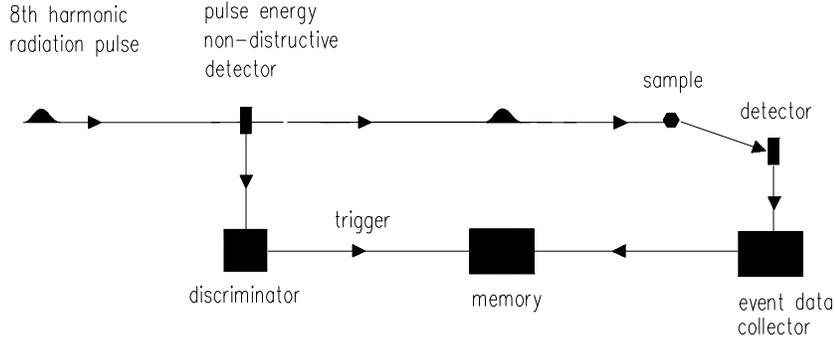,width=0.8\textwidth}
\end{center}
\caption{ Experimental setup to obtain single spike pulse duration. 
Signals from a non-destructive 8th harmonic energy pulse detector 
are used to give trigger. The energy threshold is used to reject 
events with $E$ smaller than $2\langle{E_{(8)}}\rangle$ } 
\label{fig:pp30} 
\end{figure}

\begin{figure}[tb]
\begin{center}
\epsfig{file=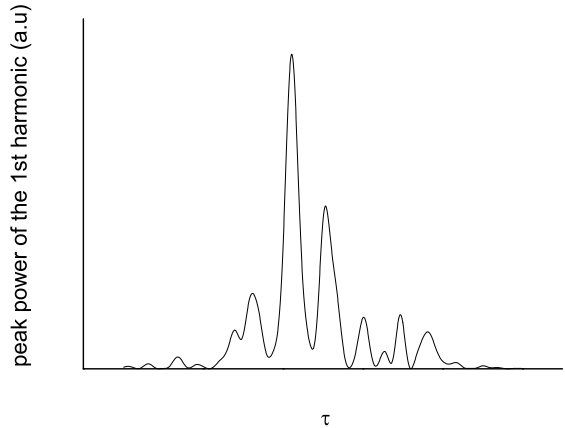,width=0.6\textwidth}
\end{center}
\caption{Illustration of the results of nonlinear transformation. 
Sample function of fundamental harmonic instantaneous power for SASE 
FEL} 
\label{fig:at1} 
\end{figure}

\begin{figure}[tb]
\begin{center}
\epsfig{file=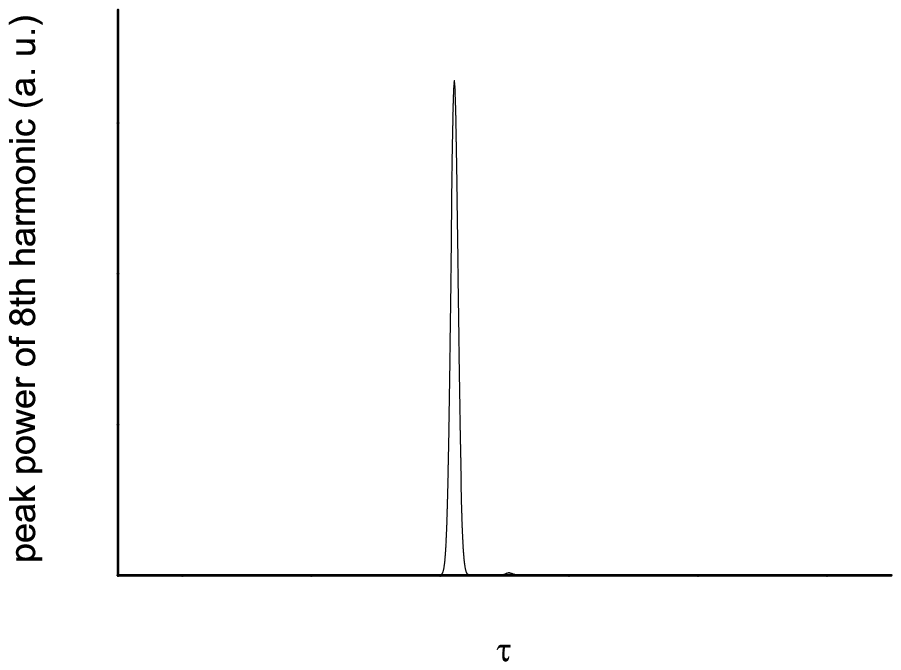,width=0.6\textwidth}
\end{center}
\caption{Illustration of the results of a nonlinear transformation. 
The nonlinear transform of Fig. (3) representing the 8th 
harmonic instantaneous power } 
\label{fig:at2} 
\end{figure}

The second development was the invention of single-bunch multistage 
High Gain Harmonic Generation (HGHG) FEL scheme \cite{sb}.  In this 
technique the second harmonic in the $n$th stage becomes the 
fundamental in the $(n+1)$th stage. Each stage (except the first one) 
consists of radiator undulator, dispersion section (demodulator), FEL 
amplifier and end-stage dispersion section (modulator).  The main 
difference with previous HGHG schemes 
\cite{ps,ps1,ps2,ps3,ps4,ps5,ps6,ps7,ps8} is that frequency 
multiplication is performed with a single electron bunch consecutively 
passing all HGHG stages.  This is possible because the density 
modulation exiting each stage is relatively small.  Hence, a small 
energy modulation is sufficient to create this microbunching in the 
dispersion section. In this case the growth of the energy spread due to 
HGHG process is much less than initial energy spread, and exponential 
growth rate in the main undulator is practically the same as without 
stage sequence. At chosen parameters for each stage the amplitude of 
the second harmonic of density modulation dominates significantly over 
the amplitude of shot noise harmonics, and the modulation of the beam 
density can be used as input signal for the next HGHG stage. 

The combination of a single-bunch multistage HGHG scheme and a 
single-spike selection technique is very promising. Our concept of 
attosecond X-ray facility is based on generation of the 8th harmonic of 
SASE radiation in the single-bunch, multistage HGHG configuration.  
Single-spike phenomena occurs when electron bunch is passed through the 
sequence of four relatively short undulators. The first stage is a 
conventional "long" wavelength (0.8 nm)  SASE FEL operating in the 
high-gain linear regime. Figure~\ref{fig:pp31} illustrates how the 
0.1~nm wavelength range may be reached by successive multiplication in 
a stage sequence (0.8 nm $\to$ 0.4 nm $\to$ 0.2 nm $\to$ 0.1 nm).

The final steps involved in obtaining a single-spike pulses of  
the 8th-harmonic radiation are as follows. The energy in the
high-harmonic radiation pulse must be measured by means of a  
non-destructive method. After each shot, 
the signal from the energy detector is sent to a discriminator having a
threshold $E_{\mathrm{th}} 
\simeq 2\langle{E_{(8)}}\rangle$, where $\langle{E_{(8)}}\rangle$ is 
the mean energy of the 8th harmonic (averaged over the ensemble of 
pulses) After discrimination signal is used to give a trigger.  A 
register is used to store information concerning the trigger and sample 
detector events.  A schematic, illustrating these processes, is shown 
in Fig.~\ref{fig:pp30}.

The possibility of  single-spike pulse selection is demonstrated in a 
simple example. With reference to Fig.~\ref{fig:at1}  consider an 
intensity function $I$ in the SASE FEL radiation pulse at fundamental 
frequency versus time $t$.  Subjecting it, for example, to a 8th 
harmonic transformation, we obtain the "image" shown in 
Fig.~\ref{fig:at2}.  An important distinction should be made between 
the sample fundamental instantaneous intensity function $I(t)$ (see 
Fig.~\ref{fig:at1}) and the transformed function $[I(t)]^{8}$ (see 
Fig.~\ref{fig:at2}).  Due to the nonlinear generation mechanism, the 
temporal structure of the 8th-harmonic radiation is similar to the 
fundamental, but with more fluctuations from spike to spike.  The fact 
that the 8th harmonic intensity is a single spike implies that the 
fluctuation of the fundamental intensity about the mean is rather 
pronounced.  How likely are we to observe a practically single bright 
spike in the intensity of the 8th harmonic radiation? Clearly, a 
necessary condition for this event is that the energy $E_{(8)}$ in the 
8th-harmonic radiation pulse is larger than the average energy 
$\langle{E_{(8)}}\rangle$. The results of numerical simulations for the 
case of the 8th-harmonic of the X-ray SASE FEL predict the probability 
of high-contrast single-spike pulses of about 1--10\% only.  For this 
method to be applicable, the electron pulse repetition rate should be 
larger than 100 pulses per second.  

This paper describes the scheme for attophysics experiments that could 
be performed at the X-ray SASE FEL at TESLA. The superconducting linear 
accelerator is an ideal accelerator to drive an attosecond XFEL. The 
high repetition rate of the TESLA accelerator (52000 pulses per second) 
should be sufficient to obtain average kHz-level pulse repetition rate 
of single spikes. The development and test of attosecond X-ray FEL at 
TESLA is greatly facilitated by the fact that undulators with required 
parameters are being developed in the framework of SASE option of X-ray 
FEL at TESLA. Also, the length foreseen for installation of SASE 
undulators is sufficient to accommodate attosecond option.

\section{Statistical properties of SASE FEL high-harmonic radiation}

The principle of operation of the proposed scheme is essentially based 
on the statistical properties of the SASE FEL harmonic radiation.  SASE 
radiation is a stochastic object and at a given time it is impossible 
to predict the amount of energy which flows to a detector.  The initial 
modulation of the electron beam is defined by the shot noise and has a 
white spectrum.  The high-gain FEL amplifier cuts and amplifies only a 
narrow frequency band of the initial spectrum $\Delta\omega/\omega \ll 
1$. In the time domain, the temporal structure of the fundamental 
harmonic radiation is chaotic with many random spikes, with a typical 
duration given by the inverse width of the spectrum envelope.  Even 
without performing numerical simulations, we can describe some general 
properties of the fundamental harmonic of the radiation from the SASE 
FEL operating in the linear regime. Indeed, in this case we deal with 
Gaussian statistics. As a result, the probability distribution of the 
instantaneous radiation intensity $I$ should be the negative 
exponential probability density distribution:  $p(I) = 
\langle{I}\rangle^{-1}\exp(-I/\langle{I}\rangle)$.  Here one should 
realize clearly that the notion of instantaneous intensity refers to a 
certain moment in time, and that the analysis must be performed over an 
ensemble of pulses. Also, the energy in the radiation pulse $E$ should 
fluctuate in accordance with the gamma distribution \cite{b}: 

\begin{displaymath}
p(E) = 
\frac{M^{M}}{\Gamma(M)}\left(\frac{E}{\langle{E}\rangle}\right)^{M-1} 
\frac{1}{\langle{E}\rangle}\exp\left(-M\frac{E}{\langle{E}\rangle}\right) \ , \end{displaymath}

\noindent where $\Gamma(M)$ is the gamma function of argument $M$, and 
$1/M = \langle{(E-\langle{E}\rangle)^{2}\rangle}/\langle{E}\rangle^{2}$ 
is the normalized dispersion of the energy distribution. These 
properties are well known in statistical optics as properties of 
completely chaotic polarized radiation \cite{g}.

Let us turn to discussion of statistical properties of the 
high-harmonic radiation in a SASE FEL. It should be noted that the 
statistics of the high-harmonic radiation from the SASE FEL changes 
significantly with respect to the fundamental harmonic (e.g., with 
respect to Gaussian statistics).  It is interesting in our case to be 
able to determine the probability density function of instantaneous 
intensity of SASE radiation  after it has been subjected to nonlinear 
transformation. We know the probability density function $p(I) = 
\langle{I}\rangle^{-1}\exp(-I/\langle{I}\rangle)$ of the fundamental 
intensity $I$, and $I$ is subjected to a transformation $z = (I)^{n}$.  
The problem is then to find the probability density function $p(z)$. It 
can be readily shown that $p(z) = 
(n\langle{I}\rangle)^{-1}{z}^{(1-n)/n}\exp(-z^{1/n}/\langle{I}\rangle)$.  
Using this distribution we get the expression for the mean value:  
$\langle{z}\rangle = n!\langle{I}\rangle^{n}$. Thus, the $n$th-harmonic 
radiation for the SASE FEL has an intensity level roughly $n!$ times 
larger than the corresponding steady-state case, but with more 
shot-to-shot fluctuations compared to the fundamental \cite{k}.  
Nontrivial behavior of the intensity of the high harmonic reflects the 
complicated nonlinear transformation of the fundamental harmonic 
statistics. One can see that Gaussian statistics is no longer valid.  

Since amplification process starts from shot noise, properties of a 
single-spike selection should be dscribed in statistical terms.  The 
statistics of concern are defined over an ensemble of radiation pulses.  
If we define the contrast $C$ as the ratio of number of photons in the 
main spike to the total number of photons in the pulse, we find that 
$\langle{C}\rangle$ asymptotically approaches unity as the ratio 
$E_{\mathrm{th}}/\langle{E_{(8)}}\rangle$ increases, where 
$E_{\mathrm{th}}$ is the threshold level of the 8th harmonic energy 
pulse discriminator.  Clearly, the larger the threshold level of 
discriminator $E_{\mathrm{th}}/\langle{E_{(8)}}\rangle$, the larger the 
number of shots per trigger pulse  $N_{\mathrm{sh}}$.  Note that the 
number of degrees of freedom $M$ of the fundamental radiation pulse is 
a parameter of the functions $\langle{C}\rangle = F(M, 
E_{\mathrm{th}}/\langle{E_{(8)}}\rangle) \ ,$ 
$\langle{N}_{\mathrm{sh}}\rangle = f(M, 
E_{\mathrm{th}}/\langle{E_{(8)}}\rangle)$ as indeed we might have 
anticipated.  

\begin{figure}[tb]
\begin{center}
\epsfig{file=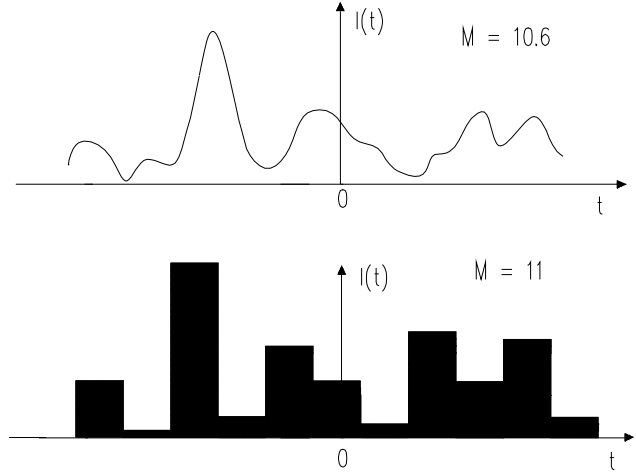,width=0.8\textwidth}
\end{center}
\caption{
Approximation of the smoothly varying instantaneous intensity by a 
"boxcar" function} 
\label{fig:pp26} 
\end{figure}

Our goal is to find the relationship between the level of the 
discriminator $E_{\mathrm{th}}/\langle{E_{(8)}}\rangle$, contrast 
factor $\langle{C}\rangle$, and number of shots per trigger pulse 
$\langle{N}_{\mathrm{sh}}\rangle$.  In addition, we wish to find 
effects of the number of modes $M$ on the single spike selection.  To 
find an approximate, density function  of integrated intensity we 
invoke a quasi-physical arguments as follows \cite{g}. As an 
approximation for fundamental harmonic, the smoothly fluctuating 
instantaneous intensity curve $I(t)$ may be replaced on the interval 
$T$ by a "boxcar" function (see Fig.~\ref{fig:pp26}). The time interval 
is divided into $M$ equal length subintervals. Within each subinterval, 
the approximation to $I(t)$ is constant; at the end of each 
subinterval, the approximate waveform jumps to a new constant value, 
assumed statistically independent of all preceding and following 
values. The probability density function of the boxcar function within 
any one subinterval is taken to be the same as the probability density 
function of the instantaneous intensity at a single time instant $t$. 

The integrated fundamental harmonic intensity is now approximated in 
terms of the area under the boxcar function:

\begin{displaymath}
E = \int I(t)\D t \simeq \sum_{i=1}^{M} I_{i}\Delta t \ ,        
\end{displaymath}

\noindent where $\Delta t$ is the width of one subinterval of the 
boxcar function and $I_{i}$ is the value of the boxcar function in the 
$i$th subinterval. By hypothesis, the probability density function of 
each $I_{i}$ is taken to be the same as the density function of the 
instantaneous intensity. Also by hypothesis, the various $I_{i}$ are 
assumed to be statistically independent.  

This particular density function of the fundamental radiation energy 
per pulse (integrated intensity) is known as a gamma probability 
density function, and accordingly the random variable $E$ is said to be 
(approximately) a gamma variate.  Continuing with the case of 
fundamental harmonic, one problem remains: the parameters of the 
density function must be chosen in such a way as to best match the 
approximate result to the true density function of $E$.  The only two 
adjustable parameters available are $\langle{I}\rangle$ and variance 
$1/M$.  The most common approach taken is to choose the parameters such 
that mean and variance of the approximate density function are exactly 
equal to the true mean and variance of fundamental radiation energy $E$ 
\cite{g}.  

It should be noted that, in a certain sense, our quasi-physical 
reasoning that led to the approximate distribution has broken down, for 
in general the parameter $M$ is not an integer, where as we implicitly 
assumed an integer number of subintervals in the boxcar function. 
In what follows we use the following assumption: $M \gg 1$. Such 
assumption does not reduce the practical applicability of the result 
obtained. It is obvious that this single-spike scheme  
has an advantage over usual SASE scheme only when the SASE radiation 
pulse is much longer than the single spike.  

Our approximate model assumes that the  smoothly fluctuating SASE 
intensity curve $I(t)$ can be replaced by a "boxcar" function.  Such a 
model allows us to  express the radiation pulse after application of 
frequency multiplication as the sum of the $M$ therms, too.  The 
integrated intensity of 8-th harmonic is now approximated in therms of 
the area under the boxcar functionals as follows:

\begin{displaymath}
E_{(8)} = \int [I(t)]^{8}\D t \simeq \sum_{i=1}^{M} [I_{i}]^{8}\Delta t 
\ .  
\end{displaymath}

\noindent Note, that $M$ is the average number of degrees  of freedom 
(or modes) in the fundamental radiation pulse.  

\begin{figure}[tb]
\begin{center}
\epsfig{file=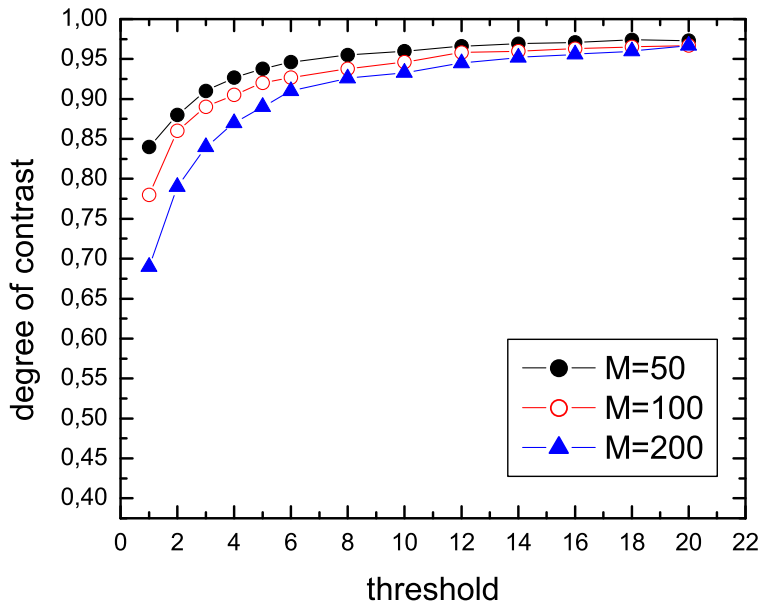,width=0.8\textwidth}
\end{center}
\caption{
Degree of contrast $\langle{C}\rangle$ versus energy threshold 
$E_{\mathrm{th}}/\langle{E_{(8)}}\rangle$ } 
\label{fig:plot2} 
\end{figure}

\begin{figure}[tb]
\begin{center}
\epsfig{file=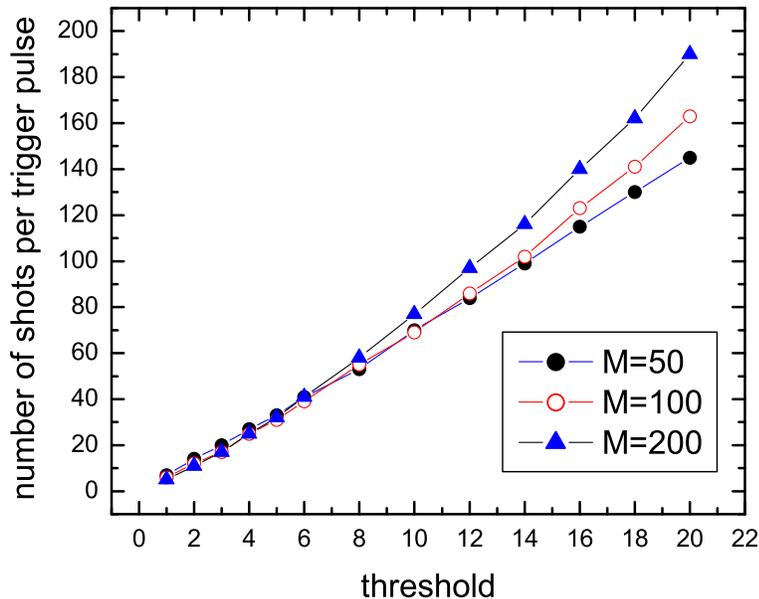,width=0.8\textwidth}
\end{center}
\caption{
Number of shots per trigger pulse 
$\langle{N_{\mathrm{sh}}}\rangle$ versus energy threshold 
$E_{\mathrm{th}}/\langle{E_{(8)}}\rangle$} 
\label{fig:plot1} 
\end{figure}

In Figs.~\ref{fig:plot2} and \ref{fig:plot1} one can see the basic 
characteristics of the single-spike pulse selection process. The 
dependence of the degree of the contrast $\langle{C}\rangle$ on the 
value of the energy threshold $E_{\mathrm{th}}/\langle{E_{(8)}}\rangle$ 
is presented in Fig.~\ref{fig:plot2}. It is seen that the contrast 
increases with an increase in the value of energy threshold, and it 
asymptotically approaches to unity. Simulations  at different values 
of $M$ show that the degree of contrast does not differ significantly 
when the number of modes is within the limits $50 < M < 200$. 
Figure~\ref{fig:plot1} shows plots of the number of shots per trigger 
pulse $\langle{N}_{\mathrm{sh}}\rangle$ versus
$E_{\mathrm{th}}/\langle{E_{(8)}}\rangle$ for several values of 
the parameter $M$. From Fig.~\ref{fig:plot1} it is quite clear that 
the dependence of $\langle{N}_{\mathrm{sh}}\rangle$ on the number of 
modes $M$ is not strong within the interval $M = 50-200$  and can  be 
ignored. 

With the preceding results in hand, it should now be possible to 
estimate, for example, the repetition rate of the single-spike pulse.  
In the case of TESLA X-ray FEL, the number of modes in the fundamental 
radiation pulse at a wavelength of 0.8 nm is about $M \simeq 50-100$.  
Suppose that we wish to achieve a contrast of 90\%.  The discriminator 
threshold required to achieve this contrast is about 
$E_{\mathrm{th}}/\langle{E_{(8)}}\rangle \simeq 2$.  If the number of 
modes is close to $M \simeq 100$, plot in Fig.~\ref{fig:plot1} shows 
that the number of shots per trigger pulse is about 10. Hence, the 
single-spike pulse repetition rate is still high (about a few thousands 
single-spike pulses per second).  On the other hand, if the contrast of 
interest is  97\% , the number of shots is about 
$\langle{N}_{\mathrm{sh}}\rangle \simeq 100$, and repetition rate of 
the single-spike pulse decreases up to a few hundreds per second.  

A few additional comments are needed in closing this section.  The 
results of numerical simulations presented above refer to the specific 
model which depends on one parameter $M$ only.  This model has proven 
to be very fruitful, providing the possibility of performing fast 
numerical simulations of the main statistical characteristics of the 
high-harmonic radiation from the SASE FEL operating in the linear 
regime. There is no doubt that these results are useful for quick 
estimate and deeper understanding of the frequency multiplication 
process in a SASE FEL. The disadvantage of this approach is in 
neglecting the radiation spectral profile and electron bunch profile 
effects. As a result, the method described above is rather crude and 
may be modified in use. The full analysis of single-spike generation, 
including both of these effects, is a very expensive for computer run.   
Fortunately, in the particular case, namely of radiation originated 
from an electron bunch with rectangular profile, a much simplified 
analysis will suffice. An approach, which takes into account the 
spectral line shape effects, uses a well known analytical 
expression for steady-state spectral Green function of the FEL 
amplifier \cite{b}. We can decompose the input 
shot noise signal into Fourier harmonics. Since in the linear regime 
all the harmonics are amplified independently, we can use the results 
of steady-state theory for each harmonic and calculate the 
corresponding Fourier harmonics of the output radiation field. An 
expression for the electric field of the electromagnetic wave as a 
function of time $t$ can be obtained using the inverse Fourier 
transform. In the framework of this model it becomes possible to 
calculate the smooth sample functions $I(t)$ and $[I(t)]^{8}$.  

To describe the single-spike selection, we should  define the degree of 
contrast.  A first question that arises is:  what is the definition of 
the main spike?  The question of when two closely spaced spikes are 
barely resolved is a complex one and lends itself to a variety of 
rather subjective answers.  One possible definition can be made as 
follows.  After analysis of smooth sample function $[I(t)]^{8}$ we find 
a time moment $t_{\mathrm{m}}$ when the intensity reaches its maximum 
value.  Then we find the number of photons within the time interval 
$(t_{\mathrm{m}} - \tau_{\mathrm{coh}}, t_{\mathrm{m}} + 
\tau_{\mathrm{coh}})$, where $\tau_{\mathrm{coh}}$ is coherence time of 
the 8th harmonic radiation pulse. In fact, the ability to resolve two 
spikes depends fundamentally on the discriminator level associated with 
the selected 8th harmonic pulse, and for this reason for large ratio, 
$E_{\mathrm{th}}/\langle{E_{(8)}}\rangle$, this problem does not exist 
at all.  It can be demonstrated that all reasonable definitions for the 
degree of contrast are consistent in the region 
$E_{\mathrm{th}}/\langle{E_{(8)}}\rangle \gg 1$.

While adding realism, the step from "boxcar" function to smooth sample 
function drastically increases computer time.  Nevertheless, one can 
learn much about accuracy of the "boxcar" scheme by seeing how it 
reproduce  "exact" results of numerical simulations for several values 
of the discriminator threshold.  Surprisingly, we can find that the 
results of the more general approach does not differ significantly with 
those of the far simpler analysis done previously.

\section{Attosecond X-ray facility description}

\begin{figure}[tb]
\begin{center}
\epsfig{file=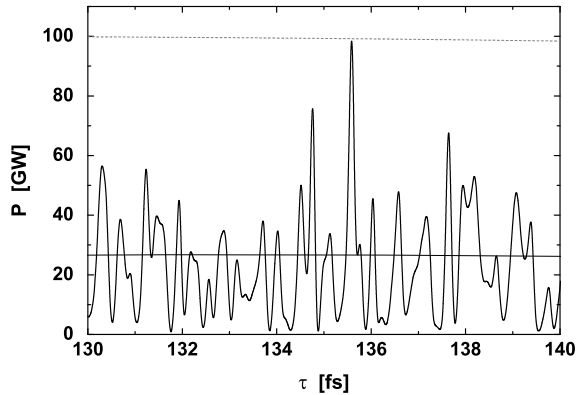,width=0.6\textwidth}
\end{center}
\caption{Temporal structure of the radiation pulse (enlarged section) 
from the SASE FEL at TESLA operating at the wavelength of 0.1 nm.  The 
solid line corresponds to averaged value.  The dashed line gives the 
axial profile of the electron bunch} 
\label{fig:p1a} 
\end{figure}

In this section we are going to discuss the application of the idea of 
the single-spike generation to a practical case, the XFEL at TESLA 
\cite{tdr}. An attosecond laser requires a broadband gain medium. What 
ultimately limits the pulse duration?  Since the temporal and spectral 
characteristics of the field are related to each other through Fourier 
transforms, the bandwidth of XFEL $\Delta\omega_{\mathrm{FEL}}$ and 
pulse duration $\tau_{\mathrm{p}}$ cannot vary independently of each 
other. There is a minimum duration-bandwidth product:  
$\Delta\omega_{\mathrm{FEL}}\tau_{\mathrm{p}} > 2\pi$.  The larger the 
FEL bandwidth is, the shorter the minimal pulse duration than can be 
obtained.  This simple physical consideration can lead directly to 
crude approximation for the minimum duration of the XFEL radiation 
pulses.  We can expect that the width of the radiation spectrum at 0.1 
nm wavelength should be of order of $\Delta\omega_{\mathrm{FEL}}/\omega 
\simeq 0.1 \%$ \cite{tdr}. Thus, the minimum duration should be 
$\tau_{\mathrm{p}} \simeq 300$ as. To illustrate the later point, let 
us consider the details of X-ray SASE FEL radiation.  The physical 
reason for this is as follows.  A SASE radiation is a situation in 
which the radiation pulse consists of a large number of independent 
wavepackets (spikes) with a typical duration given by the inverse width 
of the spectrum envelope.  Typical temporal structures of the radiation 
pulse from the X-ray SASE FEL is presented in Fig.~\ref{fig:p1a}.  The 
chaotic nature of the output radiation is a consequence of the start-up 
from shot noise. This plot also indicates that the shortest XFEL pulse 
duration is approximately $\tau_{\mathrm{p}} \simeq 300$ attoseconds.       
  
\begin{figure}[tb]
\begin{center}
\epsfig{file=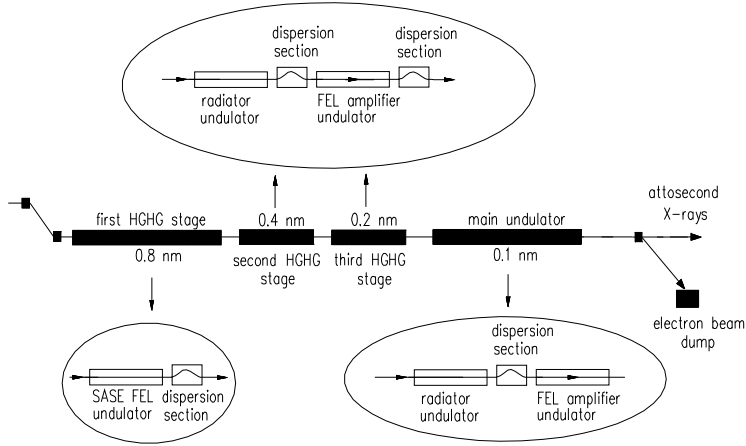,width=0.8\textwidth}
\end{center}
\caption{
Shot noise seeded single-bunch HGHG scheme proposed for attosecond 
X-ray facility } 
\label{fig:pp32} 
\end{figure}

The layout of the attosecond facility at TESLA is shown schematically 
in Fig.~\ref{fig:pp32}. Single-bunch HGHG scheme consists of three 
stages of frequency doubling and main undulator.  It has been pointed 
out that for single-spike selection scheme to work, the chain of FEL 
amplifiers must operate in linear regime. An intrinsic advantage of the 
adopted HGHG scheme is the linear mode operation. The first stage of 
frequency doubling (0.8 nm $\to$ 0.4 nm) consists of SASE FEL undulator 
and dispersion section for the beam density modulation. The scheme 
operates as follows. The first stage is a conventional FEL amplifier 
seeded by shot noise.  Radiation power is exponentially amplified upon 
passing through the first undulator.  The amplitude of energy 
modulation of the electron beam at the undulator exit is equal to 
$\delta E \simeq 0.6\Delta E_{\mathrm{in}}$, where $\Delta 
E_{\mathrm{in}}$ is the initial local energy spread in the electron 
bunch. Calculations show that in this case the beam density modulation 
is not sufficient to drive the second stage. Required value of the beam 
bunching at the second harmonic of about 5\% is achieved when the 
electron bunch passes through the dispersion section (the density 
modulation at the fundamental frequency is about 0.3).  In this 
situation the amplitude of the second harmonic of the density 
modulation dominates significantly over the amplitude of shot noise 
harmonic (of about 0.03\%), and it serves as input signal for the 
second stage.  

\begin{figure}[tb]
\begin{center}
\epsfig{file=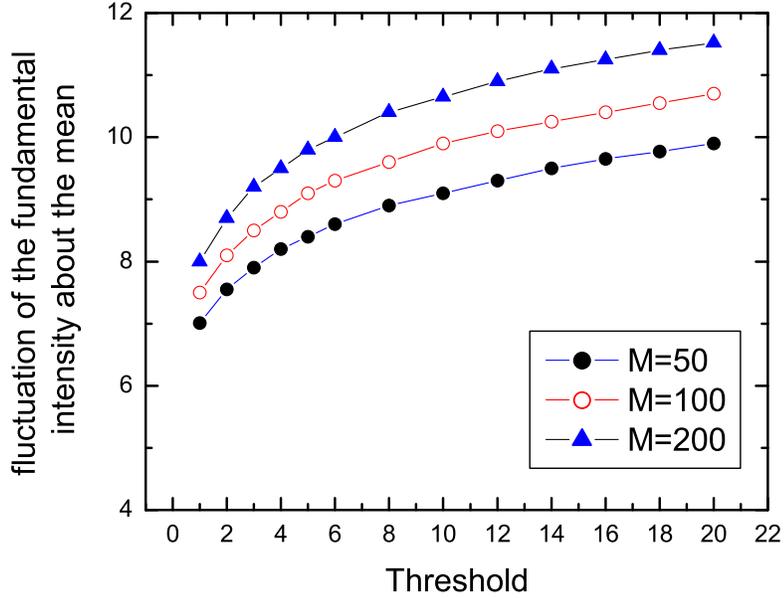,width=0.8\textwidth}
\end{center}
\caption{Fluctuation of the fundamental intensity related to 
single spike of 8th harmonic about the mean versus energy threshold } 
\label{fig:pl3} 
\end{figure}

Let us discuss the problem of optimization of the first SASE FEL. The 
function of the first SASE FEL is to prepare electron bunch with energy 
modulation at the 0.8 nm wavelength. The process of amplification of 
radiation in this undulator develops in the same way as in conventional 
SASE FEL: fluctuations of the electron beam current density serve as 
the input signal. To estimate the value of the input signal  
at fundamental harmonic related to single spike at the 8th 
harmonic, it is convenient to introduce the notion of an effective 
power of shot noise $P_{\mathrm{n}}$, which is usually used for 
numerical simulation of the SASE FEL (see, e.g., \cite{b}). For 
the 0.8 nm SASE FEL the effective shot noise power is about 
$p_{\mathrm{n}} \simeq 1$ kW.  Calculation shows (see 
Fig.~\ref{fig:pl3}) that the fluctuation of the instantaneous power at 
fundamental harmonic related to single spike at the 8th harmonic is 
about 10 times larger than the mean value.  This means that effective 
peak power of the "seed" spike is about 10 kW.  At the exit of the 
first undulator the most fraction of the electron beam has small energy 
modulation due to the FEL process except of the small ("seed" spike) 
region.  Since the instantaneous radiation power in this region is 
about 10 times larger, only this small part of the electron bunch 
produces in dispersion section the input signal at the 0.4 nm for the 
2nd HGHG stage, thus providing short pulse duration.  Upon passing 
through the first FEL undulator, radiation is exponentially amplified.  
First FEL operates in a linear regime with a power gain $G \simeq 
10^{3}$.  This value is much less than the power gain at saturation, 
$G_{\mathrm{sat}} \simeq 10^{7}$.  At such a choice of the power gain 
in the first undulator the energy modulation (in small "seed" spike 
region) induced by the FEL process amounts to 1.5 MeV and is about two 
times less than the initial energy spread (about 2.5 MeV) in the beam.  
The amplitude of the first harmonic of the beam density modulation 
related to this energy modulation is about 2\% only. Since this 
density modulation is of about 10 times smaller than required value, 
one should use dispersion section at the exit of the first stage.   

\begin{figure}[tb]
\begin{center}
\epsfig{file=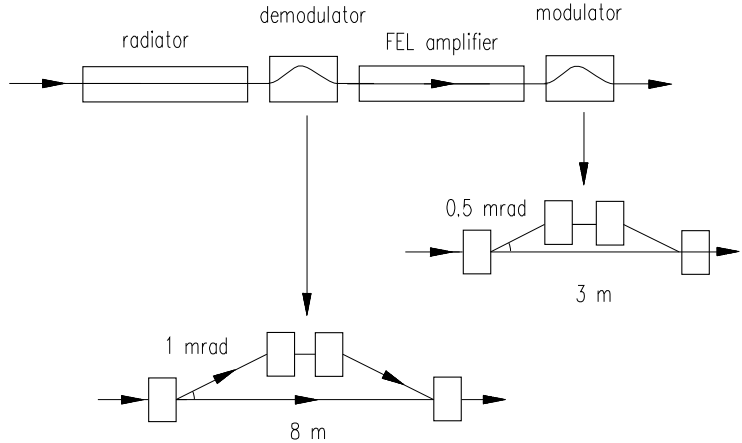,width=0.8\textwidth}
\end{center}
\caption{
Schematic illustration of design 
configuration for the second stage of  
single-bunch HGHG scheme} 
\label{fig:pp33} 
\end{figure}

The design configuration of the second stage is shown in 
Fig.~\ref{fig:pp33}. The stage consists of a short undulator 
(radiator), dispersion section (demodulator), high-gain FEL amplifier 
and end-stage dispersion section (modulator). Following the first stage 
the beam and seed radiation enter short undulator (radiator) which is 
resonant with the second harmonic of the seed radiation (0.4 nm). In 
the radiator the seed radiation at fundamental harmonic plays no role 
and is diffracted out of the electron beam.  However, a new 0.4 nm 
radiation is generated by the density-modulated electron beam and 
rapidly reaches 10 MW-level peak power.  After the radiator the 
electron beam is guided through a dispersion section (magnetic 
chicane). The trajectory of the electron beam in the chicane has the 
shape of an isosceles triangle with the base equal to $L$. The angle 
adjacent to the base, $\theta$, is considered to be small.  The problem 
of suppressing the beam modulation induced in the first stage can be 
solved quite naturally due to the presence of the local energy spread 
in the electron beam.  Parameters in our case are:  $\theta = 1$ mrad, 
$L = 8$ m, compaction factor $R_{56} = L\theta^{2} \simeq 8 \mu$m, 
$\sqrt{\langle(\Delta E)^{2}\rangle}R_{56}/E_{0} \simeq$ 0.8 nm, 
$\lambda$ = 0.8 nm. This leads to the suppression of the electron beam 
modulation by a factor $\exp(- 20)$ (for Gaussian energy distribution).  
The demodulators needed for HGHG stages has to satisfy two additional 
requirements. First, the radiation pulse must overlap the electron 
bunch at the chicane exit, i.e. the electron beam extra path length 
must be much smaller comparing with the electron bunch length. Second, 
coherent synchrotron radiation (CSR) effects should be avoided in order 
tp preserve transverse emittance.  In the present design of HGHG scheme 
we have only about 4~$\mu$m extra path length for the electron beam, 
while the rms length of electron bunch is about 25 $\mu$m, thus delay 
effect is negligible.  Calculation of the CSR effects shows that this 
should not be a serious limitation in our case.         

Passing the chicane the demodulated electron beam and seed 0.4 nm 
radiation enter the FEL amplifier undulator.  This undulator is long 
enough to reach  0.6 $\Delta E_{\mathrm{in}}$ energy modulation at the 
wavelength of 0.4 nm.  Since the density modulation at 0.4 nm is of 
about 10 times smaller than required value, one should use dispersion 
section at the exit of the second stage.  After passing second 
dispersion section, the energy modulation induced in the beam by 
amplification in the FEL amplifier transforms into the density 
modulation. The values of the second (0.4 nm) and the 4th (0.2 nm) 
harmonics of density modulation at the second stage exit are about 30\% 
and 5\%, respectively. These values are approximately the same as the 
amplitudes of the first and second harmonics at first stage exit. 
				       
Following the second stage the beam enters the third 
stage which is resonant with the 4th harmonic (0.2 nm). Like the second 
stage, the third stage also consists of the radiator, demodulator, FEL 
amplifier and modulator.  Now 0.2 nm density modulation serves as a 
seed for this radiator. The dispersion section needed for third stage 
has only about 2 $\mu$m extra path length for the electron beam. In 
this case no problem of synchronization between radiation pulse and 
electron bunch occurs. The length of the FEL amplifier undulator is 
chosen in such a way that the energy modulation at the undulator exit 
has the value of $\delta E \simeq \Delta E_{\mathrm{in}}$.  Required 
value of the beam bunching at the 8th harmonic (0.1 nm) of about 10\% 
is achieved when the electron beam passes through dispersion section.   

Finally, after the third stage the electron beam enters the main 
undulator system which is resonant to the 8th harmonic (0.1 nm). This 
undulator system consists of the radiator undulator, dispersion section 
and FEL amplifier which operates  in a linear regime with a power gain 
of about $G \simeq 10^{2}$.  This value is much less than the power 
gain of 0.1 nm SASE FEL at saturation, $G_{\mathrm{sat}} \simeq 
10^{6}$.  The attosecond XFEL will provide transversely coherent 
bandwidth limited single-spike pulses.  Frequency multiplication can be 
an essential pulse shortening mechanism in a multistage HGHG scheme.  
Because of the nonlinearity of the conversion process, one expects the 
generation of double frequency pulse $\sqrt{2}$ times shorter than the 
pulse duration at the fundamental frequency (for Gaussian profile).  
Successive multiplication to the $n$th harmonic resulting in 
$\sqrt{n}$-fold compression of the $n$th harmonic pulse duration. In 
our case $n = 8$, and numerical simulations show that we can obtain 
600~attosecond pulses at the wavelength of 0.1 nm. The number of 
photons can exceed $10^{10}$ per pulse.

\begin{figure}[tb]
\begin{center}
\epsfig{file=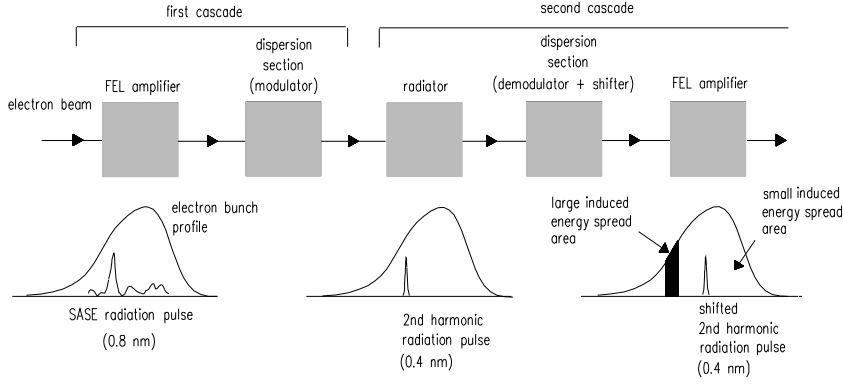,width=0.8\textwidth}
\end{center}
\caption{
Sketch of the second harmonic generation from the first to the second 
HGHG cascade} 
\label{fig:pp27a} 
\end{figure}

The small energy perturbation of the electron beam is one of the 
advantages of the adopted HGHG FEL design.  In our scheme the energy 
modulation (i.e.  correlated) energy spread induced in the $n$th stage, 
transforms to local (i.e.  uncorrelated) energy spread in the $(n+1)$th 
stage. As a result, the dispersion of the electron energy distribution 
at the exit of the multistage scheme is calculated as the sum of 
induced dispersions. For instance, the total energy spread generated to 
the end of the first demodulator, can be estimated as

\begin{displaymath}
\sqrt{\langle(\Delta E)^{2}\rangle} \simeq \sqrt{(\Delta 
E_{\mathrm{in}})^{2} + 0.5\times (\delta E)^{2}} \simeq 1.01 
\Delta E_{\mathrm{in}} 
\end{displaymath}                         
  
\noindent for  $\delta E \simeq 0.2 \Delta E_{\mathrm{in}}$.  Such a 
small degradation of the energy spread almost does not lead to 
degradation of output radiation.

One can wonder why presenting calculation of the energy spread induced 
in the first stage we use the value $\delta E \simeq 0.2 \Delta 
E_{\mathrm{in}}$. It should be reemphasized that the function of the 
first SASE FEL is to prepare electron bunch with $\delta E \simeq 0.6 
\Delta E_{\mathrm{in}}$ energy modulation at the 0.8 nm wavelength.  
Here we direct the attention of the reader to the fact that this level 
of the energy modulation takes place only at that part of the electron 
bunch, defined by the length of the "seed" spike.  Fluctuation of the 
instantaneous power at fundamental harmonic related to single spike at 
the 8th harmonic is about 10 times larger than mean value (see 
Fig.~\ref{fig:pl3}).  This means that at the exit of the first 
undulator the most fraction of the electron beam has 3 times smaller 
energy modulation due to the FEL process except of the small ("seed" 
spike) region. The resulting radiation produced in the radiator of the 
second stage is shifted to the part of electron bunch with small 
induced energy spread. This shift is done by passing the electron beam 
through the first magnetic chicane (demodulator)  so as to delay the 
electrons by a time equal to 12 fs. On the other hand, the single-spike 
pulse duration at 0.8 nm wavelength is about 2 fs only. A schematic 
diagram of the second harmonic generation is shown in 
Fig.~\ref{fig:pp27a} as well as the relative positions of the radiation 
and electrons before and after the bunch shift. This is to some degree 
surprising:  best results were obtained in hybrid operation, using 
chicane as demodulator, and as magnetic delay. So, this is the 
important concept:  we use a magnetic delay to position the 0.4 nm 
radiation at the "fresh" part of the electron bunch. The radiation 
starts interacting with a new set of electrons, which have the small 
energy spread, since they did not participate in the previous 
interaction with "seed" spike. Actually, this is the essence of the 
"fresh bunch" techniques which was introduced in \cite{fb}. The ability 
to shift the radiation to "undisturbed" part of the electron bunch 
after each harmonic generation is important advantage of the proposed 
scheme.  This advantage (hybrid operation) explains why we expect a 
total energy spread growth at the main (0.1 nm) undulator of about a 
few per cent only.  

\begin{table}[p]

\caption{Parameters of the gap-tunable undulators at TESLA}
\medskip

\begin{tabular}{| l l l |}
\hline
&
SASE4&
SASE5\\
\hline
\hspace*{5pt} Type & planar& circular \\ 
\hspace*{5pt} external beta-function, m           & 45& 45  \\ 
\hspace*{5pt} magnetic period, mm     & 60& 107 \\ 
\hspace*{5pt} electron energy, GeV     & 25& 25 \\ 
\hspace*{5pt} magnetic gap, mm     & 22-12& 35-12 \\ 
\hspace*{5pt} magnetic field, T    & 0.66-1.33& 0.38-0.96 \\ 
\hspace*{5pt} photon wavelength, nm    & 0.1-0.35& 0.4-2.5  \\ 
\hspace*{5pt} parameter $\rho$            & 4-6 $[10^{-4}]$  & 15-30 
$[10^{-4}]$ \\ 
\hspace*{5pt} total length, m           & 323& 177  \\ 
\hline 
\end{tabular} 
\label{t1} 
\end{table}                                        

Let us to discuss some general aspects of noise influence on the 
attosecond facility operation. In \cite{sb} we performed a detailed 
study of process of amplification in the HGHG FEL schemes taking into 
account shot noise in the electron beam. It has been found that a 
general disadvantage of HGHG FEL schemes (as well as any frequency 
multiplication scheme) is strong noise degradation of the properties of 
output radiation with increasing harmonic number $n$. In the case of 
HGHG FEL this means that the effect of frequency multiplication by a 
factor of $n$ results in multiplication of the ratio of noise power to 
carrier $(P_{\mathrm{s}}/P_{\mathrm{n}})$  by a factor of $n^2$.  An 
attractive feature of proposed HGHG scheme is that multiplication 
factor is equal to $n = 8$ only. Here it is relevant to remember that 
amplification process in attosecond facility starts from shot noise and 
problem of noise in the first HGHG cascade does not exist at all. 
Considering the other contributions to the noise output, it is obvious 
that the condition $(P_{\mathrm{s}}/P_{\mathrm{n}}) \gg 1$ will not be 
violated if when taking into account of the second and third stage 
contributions to the noise output.  It is important to note that noise 
degradation discussed above takes place at that part of the 
electron bunch, defined by the length of the "seed" single spike.  
Nevertheless, the problem of noise degradation is more complicated and 
there is another noise effect that can be important for ultra-short 
X-ray pulse generation. We must account to that shot noise in the main 
part of electron bunch can provide long pulse SASE radiation in the 
main undulator. This effect leads to degradation of contrast of output 
attosecond X-ray pulses.  However, parameters of the third stage and 
main undulator can be optimized in such a way that this degradation of 
contrast is insignificant. 

Now we would like to discuss in more detail a single-spike contrast 
preservation during amplification process in the main undulator.  The 
long main  undulator can emit a SASE radiation pulse with duration of 
about 200 times longer than the seed ultra-short pulse. In view of 
this, the preservation of degree of single-spike pulse contrast is 
clearly of critical importance to the operation and scientific utility 
of the attosecond X-ray facility. The criteria that led to the 
selection of the third cascade and main undulator parameters were 
established by first determining the minimum value of main undulator 
gain. The smaller the main undulator gain, the better the contrast of 
attosecond pulses and additionally the smaller the cost of undulator 
systems. For this reason, the best way to reduce the gain of main 
undulator is by generating maximum 8th harmonic in the spatial bunching 
at the third cascade exit. The optimum spatial bunching, keeping the 
linear mode operation, results in an amplitude of 8th harmonic of 
10\%.  Optimized gain of the main undulator is equal to $G \simeq 
10^{2}$. Calculation shows that in this case the ratio of the SASE 
pulse energy and attosecond pulse energy at the main undulator exit 
reaches a value of about per cent only. Thus, we find that effects of 
SASE radiation in the main undulator are not important in our case.   

For the TESLA XFEL project five SASE FELs are planned in 
total \cite{tdr}. Two undulators (SASE2, SASE3) have a fixed gap. On 
SASE1, SASE4, SASE5 undulators the magnet gap will be varied 
mechanically for wavelength tuning. The total length of 
undulator systems is 1700~m.  This huge length, with as much as 281 
segments, calls for standardization and an economic design, that is 
optimized for the production of large quantities.  The total length of 
an undulator system is much longer than the optimum $\beta$-function.  
External strong focusing is therefore needed to keep the 
$\beta$-function within limits acceptable for the FEL process.  The 
undulator system may be separated into undulator segments and strong 
focusing quadrupoles. The undulator system for an X-ray FEL is a long 
periodic array of undulator segments and intersections.  Each cell 
consists of one undulator segment and the components in the 
intersections such as a phase shifter, correctors, a quadrupole, and 
beam position monitors. Table \ref{t1} summarizes the proposed 
design parameters for gap-tunable SASE4 and SASE5 
undulators.  We mention it here only for illustration. The HGHG scheme 
also requires the use of gap tuning. The great advantage of proposed 
design is that attosecond X-ray facility undulators are shortened 
versions of the SASE4 and SASE5 undulators.  The proposal for the 
undulator systems for HGHG scheme is based exclusively on the 
standardized components. Only two different types of magnet structures 
are needed.  The total magnetic length of SASE5-type structure for 
first and second HGHG stages is 80~m, and length of SASE4-type 
structure for the third stage and main undulator is 160~m. This length 
is taken as the net magnetic length of the undulator. The total length 
of undulator systems includes 20\% contingency for field errors, 
misalignment, etc.   

\section*{Acknowledgments}

   We thank 
W.~Brefeld, 
B.~Faatz, 
J.~Feldhaus, 
M.~K\"orfer, 
J.~Krzywinski, 
T.~M\"oller, 
J.~Pfl\"uger, 
J.~Rossbach,  
and S.~Schreiber 
for many useful discussions. 
We thank J.R.~Schneider and D.~Trines for their interest in this work 
and support.


\clearpage

\begin{thebibliography}{99}
  


\bibitem{ay}
V.~Ayvazyan et al., DESY-print 01-226(2001), Phys. 
Rev. Lett. 88(2002)104802

\bibitem{tdr}
F. Richard et al. (eds), TESLA Technical Design Report, DESY2001-011, 
and http://tesla.desy.de


\bibitem{ko}
M. Koerfer, "The TTF-FEL status and its future as soft X-ray user 
facility" presented at 23rd Int. FEL Conf. (Darmstadt, Germany 2001), 
Nucl. Instrum. and Methods A in press.


\bibitem{tr}
W. Brefeld, et al., Preprint DESY 02-038, DESY Hamburg, 2002 


\bibitem{sb}
E. L. Saldin, E. A. Schneidmiller and M. V. Yurkov, 
Opt. Commun. 202(2002)169

\bibitem{ps}
I. Boscolo, V. Stagno, Nuovo Cimento B 58(1980)267

\bibitem{ps1}
I. Schnitzer, A. Gover, Nucl. Instr. Meth. A 237(1985)124

\bibitem{ps2} 
R. Bonifacio, L. De Salvo, P. Pierini, Nucl. Instr. Meth. A 
293(1990)627

\bibitem{ps3}
L. H. Yu, Phys. Rev. A 44(1991)5178

\bibitem{ps4}
I. Ben-Zvi et al., Nucl. Instr. Meth. A 304(1991)151

\bibitem{ps5}
I. Ben-Zvi et al., Nucl. Instr. Meth. A 393(1997)II-10

\bibitem{ps6}
L. H. Yu, I. Ben-Zvi et al., Nucl. Instr. Meth. A 393(1997)96

\bibitem{ps7}
L. H. Yu et al., Nucl. Instr. Meth. A 445(2000)301

\bibitem{ps8}
J. Wu, L. H. Yu Nucl. Instr. Meth. A 475(2001)104

\bibitem{fb}
I. Ben-Zvi, K. M. Yang, L. H. Yu, Nucl. Instr. Meth. A 318(1992)726


\bibitem{b}
E. L. Saldin, E. A. Schneidmiller and M. V. Yurkov, 
"The Physics of Free Electron Lasers" (Springer-Verlag, 
Berlin-Heidelberg-New York, 1999)  


\bibitem{g}
J. Goodman, "Statistical Optics" (Willey, New York, 1985) 

\bibitem{k}
Z. Huang and K.-J. Kim, Phys. Rev. E 62, 7295(2000)

\end{thebibliography}
\end{document}